\documentclass{aa}

\usepackage{psfig,graphicx,epsfig}
\usepackage{lscape}

\newcommand{\rx}{RX J0852.0-4622}             
\newcommand{\cx}{CXOU J085201.4-461753}       

\begin{document}

 \title{Exploring the Central Compact Object in the \rx\ Supernova Remnant with XMM-Newton}

 \titlerunning{Exploring the Central Compact Object in \rx}
 \authorrunning{W.Becker et al.}

 \author{W. Becker\inst{1} \and C. Y. Hui\inst{1} \and B. Aschenbach\inst{1} \and A. Iyudin\inst{2}}
 \offprints{W.Becker (web@mpe.mpg.de)}
 \institute{Max-Planck Institut f\"ur Extraterrestrische Physik, Giessebachstrasse 1, D-85741 Garching, Germany \and 
 Skobeltsyn Institute of Nuclear Physics, Moscow State University, Vorob'evy Gory, 119992 Moscow, Russia}
 
 \date{Submitted to A\&A on July 5, 2006}

 \abstract{The properties of the presumably young galactic supernova remnant (SNR) RX 
 J0852.0-4622,  discovered by ROSAT, are still uncertain. The data concerning the distance 
 to the SNR, its age, and the presence of a compact remnant remain controversial. We report 
 the results of several XMM-Newton observations of CXOU J085201.4-461753, the central 
 compact source in RX J0852.0-4622. The currently prefered interpretation of 
 CXOU J085201.4-461753 being a neutron star is in line with our analysis. The Chandra 
 candidate pulsation periods are not confirmed; actually no period was found down to 
 a $3 \sigma$ upper limit for any pulsed fraction. The spectrum of CXOU J085201.4-461753
 is best described by either a two blackbody spectrum or a single blackbody 
 spectrum with a high energy power law tail. The two blackbody temperatures of 4 MK and 6.6 MK 
 along with the small size of the emitting regions with radii of 0.36 and 0.06 km invalidate 
 the interpretation that the thermal radiation is cooling emission from the entire neutron 
 star surface. The double blackbody model suggests emission from the neutron star's hot polar 
 regions. No X-ray lines, including the emission feature previously claimed to be present in 
 Chandra data, were found.  

 }
   
 \maketitle


\section{Introduction}
 A very fascinating result reflecting the advantage of wide-field spectro-imaging
 observations provided by ROSAT in its all-sky survey (RASS) is the discovery
 of the supernova remnant \rx\ by  Aschenbach (1998). The remnant is located  
 along the line of sight towards the south-east corner of the Vela supernova  
 remnant (cf.~Fig.\ref{Vela_Jr_Rosat}). Because the latter dominates the emission 
 in the soft band, \rx\ is best visible in the RASS data at energies above 
 $\sim 1$ keV. The remnant has a circular shell-like shape with a diameter of 
 2 degrees. Although the distance to the remnant is uncertain, its 
 free-expansion age ($t\sim 3.4 \times 10^3 v_{5k}^{-1} d_{1kpc}\,\mbox{yr}$, 
 with $v_{5k}$ as expansion velocity in units of 5000 km/s) suggests a remnant
 age much younger than that of Vela. Because of its suggested age, it is often 
 denoted as {\em Vela-Junior}. 

 COMPTEL data suggest that the remnant has a 1.157 MeV  $\gamma$-line from the 
 radioactive isotope $^{44}$Ti associated with it (Iyudin et al.~1998). Together 
 with ROSAT X-ray data, Aschenbach et al.~(1999) suggested the remnant could be 
 as young as $\sim 680$ years and as close as $\sim 200$ pc, which would make it 
 the closest and youngest supernova remnant among the $\sim 231$ galactic remnants known 
 today (Green 2004).

 However, reanalysis of COMPTEL data indicates that the detection of the $^{44}$Ti
 line in the direction of \rx\ is only significant at the level of $\sim (2-4)\sigma$  
 (Sch\"onfelder et al. 2000). 
 Discarding the $^{44}$Ti line detection would 
 leave the age and distance estimates significantly less constrained. 
 ASCA observations have shown that the hydrogen 
 column absorption of the remnant is about one order of magnitude larger than that 
 of the Vela supernova remnant (Tsunemi et al.~2000; Slane et al.~2001) which would
 suggest that \rx\ is far behind Vela. On the other hand, the 
 lack of strong variation in the column absorption across  \rx\ further indicates 
 that the remnant cannot be more distant than the Vela Molecular Ridge,  
 i.e. 1 to 2 kpc (Slane et al.~2001).

 The X-ray spectra from \rx\ as obtained by the ASCA GIS detector are well described 
 by a power law model which indicates that the X-ray emission of \rx\ is likely to be 
 predominately non-thermal (Tsunemi et al.~2000; Slane et al.~2001). Together with 
 SN1006 and G347.3-0.5, 
 \rx\ thus forms the exclusive group of non-thermal shell-type supernova remnants which 
 are believed to be accelerators of cosmic rays. This is further supported by the recent 
 detection of TeV $\gamma -$ray images  by HESS which correlate well with the 
 X-ray images (Aharonian et al.~2005). 

 Although the GIS spectra appear to be featureless and power law like, analysis of the 
 spectrum of the north-western rim obtained by the ASCA SIS detector has shown a line 
 feature at $\sim 4.1$ keV, though its significance was low (Tsunemi et al.~2000; 
 Slane et al.~2001). The recent study of the rim of \rx\ with XMM-Newton has confirmed the 
 non-thermal nature of the remnant as well as the presence of an emission line feature at 
 4.45$\pm$0.05 keV (Iyudin et al.~2005). The authors suggest that the line (or lines) originate 
 from the emission of Ti and Sc which might be excited by atom/ion or ion/ion high velocity 
 collisions. This is in agreement with the width of the 1.157 MeV $\gamma -$ray line which 
 indicates a large velocity of the emitting matter of $\sim 15000$ km/s 
 (Iyudin et al.~1998). Furthermore, the 
 consistency of the X-ray line flux and the $\gamma$-ray line flux support the 
 existence and the amount of Ti in \rx\ claimed by Iyudin et al.~(1998). 
 However, if the interpretation
 of Iyudin et al.~(2005) is correct, only sub-Chandrasekhar type Ia supernova can produce such
 high ejecta velocity within the current knowledge of explosion model. On the other hand 
 a sub-Chandrasekhar type Ia supernova would not leave a compact stellar-like object, in contrast 
 to the observations. Apart from the 1.157 MeV 
 line, the decay chain $^{44}$Ti$\rightarrow^{44}$Sc$\rightarrow^{44}$Ca also produces  
 the hard X-ray lines at 67.9 keV and 78.4 keV. So far INTEGRAL observations have not found 
 any evidence of these two lines (Renaud et al.~2006). 

 \begin{center}
 \begin{figure}[h]
 \centerline{\psfig{file=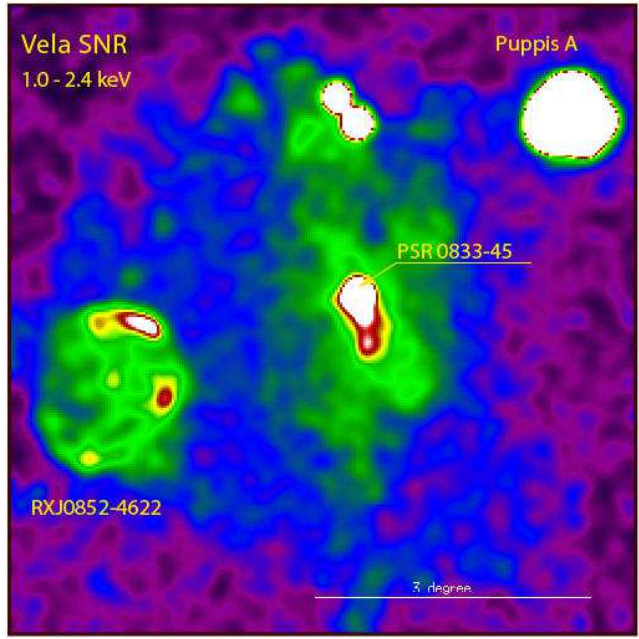,width=8.45cm,clip=}}
 \caption[]{\renewcommand{\baselinestretch}{1.0}\small\normalsize\small
  {\em Vela and friends}: The two supernova remnants Puppis-A and \rx\ are
  both located at the edge of the Vela supernova remnant. Emission from the 
  central compact source \cx\ is clearly visible in the ROSAT 
  all-sky survey image near to the center of \rx.} \label{Vela_Jr_Rosat}
  \end{figure}
  \vspace{-1cm}
  \end{center}

 A central compact X-ray source \cx\ in \rx\ was first recognized as a $5\sigma$ 
 excess in the RASS data by Aschenbach et al.~(1998). Its existence was finally 
 confirmed with the detection of a hard, somewhat extended X-ray source in the ASCA 
 data (Tsunemi et al.~2000; Slane et al.~2001). To conclusively identify this source 
 as the compact remnant of the supernova explosion that created \rx\ was not possible 
 because of the presence of two foreground stars (HD 76060 and Wray 16-30) which could 
 be responsible for at least part of the central X-ray emission. On the other hand, 
 BeppoSAX observations have revealed another hard X-ray source, SAX J0852.0$-$4615, which 
 was not resolved in previous observations (Mereghetti 2001). Based on its harder 
 X-ray spectrum and higher X-ray to optical flux ratio, the authors found SAX J0852.0$-$4615 
 to be more likely a neutron star candidate associated with \rx. 

\begin{table*}
 \centering
 \caption{Details of XMM-Newton observations of \cx. In the April 2001 observation \cx\ got
 placed right on a PN CCD gap. We therefore don't give a counting rate.}
 \label{obs_sum}
 \begin{tabular}{c c c c c c}
 \\
 \hline\hline
 Detector     &  Instrument Mode    & Filter   & Start Date & Effective & Net Rate\\
              &                     &          &            & Exposure  & (cts $\mbox{s}^{-1}$)\\
 \hline\\[-2ex]
 \multicolumn{6}{c}{\bf Obs. ID: 0112870601}\\\\[-2ex]
 \hline
   MOS1       & Full Frame          & Medium   & 2001-04-27  & 8 ksec & $0.146\pm0.004$ \\
   MOS2       & Full Frame          & Medium   & 2001-04-27  & 8 ksec & $0.149\pm0.004$ \\
   PN         & Extended Full Frame & Medium   & 2001-04-27  & 6 ksec &         {}      \\
 \hline\\[-2ex]
 \multicolumn{6}{c}{\bf Obs. ID: 0147750101}\\\\[-2ex]
 \hline
   MOS1       & Full Frame          & Medium   & 2003-05-21  & 38 ksec & $0.155\pm0.002$ \\
   MOS2       & Full Frame          & Medium   & 2003-05-21  & 38 ksec & $0.156\pm0.002$ \\
   PN         & Small Window        & Thin     & 2003-05-21  & 25 ksec & $0.477\pm0.005$ \\
 \hline\\[-2ex]
 \multicolumn{6}{c}{\bf Obs. ID: 0147750201}\\\\[-2ex]
 \hline
   MOS1       & Full Frame          & Medium   & 2003-06-25  & 12 ksec & $0.156\pm0.004$ \\
   MOS2       & Full Frame          & Medium   & 2003-06-25  & 12 ksec & $0.154\pm0.004$ \\
   PN         & Small Window        & Thin     & 2003-06-25  & 8 ksec  & $0.472\pm0.008$ \\
 \hline
 \end{tabular}
 \end{table*}

 A first Chandra observation of the central part of Vela-Jr was performed in October, 2000.
 The superior angular resolution of Chandra clearly resolved a point-like source, 
 \cx, and invalidated the association with the aforementioned foreground stars 
 (Pavlov et al.~2001). Moreover, there was no source detected at the position of 
 SAX J0852.0$-$4615. In view of this, there is little doubt that \cx, located only 
 $\sim 4$ arcmin off the geometric center of \rx, is the compact remnant 
 of the supernova explosion. This is further suggested by the $f_x/f_{opt}$ ratio as 
 the source has no optical counterpart known brighter than $m_B > 22.5$, $m_R > 21.0$ 
 (Pavlov et al.~2001).  

 The first short 3 ksec Chandra observation provided only an accurate 
 position for \cx. Spectral analysis with this data were strongly hampered by 
 low photon statistics and pile-up effect (Pavlov et al.~2001). A second 
 31.5 ksec Chandra observation performed in August, 2001, aimed for timing and 
 spectral analysis and was set up in ACIS continuous clocking mode which provides
 2.85 ms temporal resolution at the expense of spatial information in one dimension.
 In this data it was found that the spectrum of \cx\ can be well modeled by a single 
 blackbody with a temperature of $\sim 4.68\times 10^{6}$ K and a projected emitting 
 area with a blackbody radius of $\sim 280$ m (Kargaltsev et al.~2002). From a timing 
 analysis these authors reported two candidate periods of $\sim 301$ and $\sim 33$ ms,  
 though the significance of these signals was rather low. Furthermore, no evidence 
 for any long-term variability of  the source flux was found (Kargaltsev et al.~2002). 

 Although there is no indication of diffuse nebular emission around the point source in 
 X-rays (Pavlov et al.~2001; Kargaltsev et al.~2002), Pellizzoni, Mereghetti, \& De Luca 
 (2002) detected a faint and circular $H_{\alpha}$ nebula with a diameter of $\sim 6$ arcsec 
 at the position of \cx. These authors suggest that the nebula emission could be created   
 by a bow shock which is driven by the relativistic wind of a neutron star.
 
 Many previous studies favor the interpretation of \cx\ being a neutron star. It should be 
 admitted, though, that the properties of this compact object are not tightly constrained 
 and hence its nature cannot be resolved without ambiguity. It is further still on discussion 
 whether this source is physically associated with \rx. Answering this question can help us 
 to constrain the nature of \rx\ in turn. 

 In this paper, we present a first detailed study of \cx\ and its environment by making use
 all XMM-Newton data taken before 2005\footnote{Tentative results on the analysis of parts 
 of this data were already published by Becker \& Aschenbach (2002).}.In \S2, we give a brief 
 description of the performed observations. In \S3, we present the methods and results of 
 our data analysis which are subsequently discussed in \S4.

\section{Observations \& Data Reduction}

 By early 2005 three observations have been targeted with XMM-Newton on \cx. The basic 
 information of the observations are summarized in Table 1. The huge collecting 
 power and high spectral resolution of XMM-Newton provides us with data of much 
 higher photon statistics than it was achieved with Chandra and other previous observatories. 
 
 \cx\ was first observed by XMM-Newton on April 27, 2001 for a duration of $\sim 25$ ksec. 
 The MOS1/2 and the PN cameras were operated in full-frame and extended full-frame mode, 
 respectively. The medium filter was used in all instruments to block stray light and 
 optical leakage from bright foreground stars. Since the accurate position of the central 
 compact object was not known at the time of this first XMM-Newton observation, \cx\ got 
 placed 2.3 arcmin off-axis. The point spread function at this off-axis angle 
 is not much different from the on-axis one, but \cx\ got located right on a CCD 
 gap in the PN-camera, causing a severe decrease in photon statistics and data quality.
 We therefore excluded this PN dataset from our analysis. 

 The second and third XMM-Newton observations of \cx\ were performed on May 21, 2003  and 
 June 25, 2003 with total exposure times of $\sim43$ and $\sim17$ ksec, respectively. 
 \cx\ was observed on-axis in both observations. The MOS1/2 cameras were setup in the 
 same way as in the 2001 observation. The EPIC-PN camera was operated in small-window 
 mode with a thin filter. 

 All the raw data were processed with version 6.1.0 of the XMM Science Analysis Software 
 (XMMSAS). We created filtered event files for the energy range $0.3- 10$ keV and selected
 only those events for which the pattern was between $0-12$ for the MOS cameras and $0-4$ for 
 the PN camera. We further cleaned the data by accepting only good time intervals (when 
 the sky background was low) and removed all events potentially contaminated by bad pixels. 
 The effective exposure times after data cleaning and the background corrected
 source count rates in the $0.3-10$ keV band are summarized in Table 1. 

 \section{Data Analysis}

 \subsection{Spatial Analysis}

 \begin{figure}[t]
 \centerline{\psfig{file=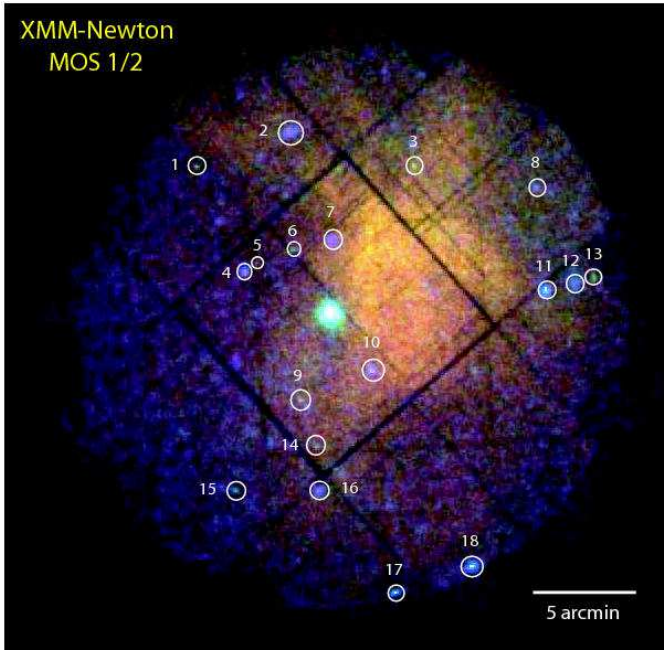,width=8.8cm,clip=}}
 \caption[]{\renewcommand{\baselinestretch}{1.0}\small\normalsize\small
  XMM-Newton MOS1/2 false color image of the inner 30 arcmin central region of
  \rx\ (red: $0.3-0.75$ keV; green: $0.75-2$ keV and blue: $2-10$ keV). The central bright
  source is \cx. 18 other point-like sources, marked by circles,
  are detected in the field of view (cf.~Table \ref{sources}).
  The binning factor of this image is 4 arcsec.
  %
  %
  } \label{mos12_image}
 \end{figure}

 A false color image of the merged MOS1/2 data from all three XMM-Newton observations 
 is displayed in Figure \ref{mos12_image}. \cx\ is the brightest source in the field. 
 It is located at the western edge of a faint $\sim 9\times 14$ arcmin wide diffuse 
 and irregular shaped extended X-ray structure which emits mostly in the soft bands 
 below $\sim 2$ keV. The extended emission was already visible in ASCA images, but 
 less resolved from and less discriminating the compact source because of ASCA's 
 wide point spread function. 
 The XMM-Newton observations provide for the first time a more detailed view of 
 this field. 
  
 Apart from \cx\ we have detected 18 point-like sources serendipitously in the field 
 of view. Their positions are given in Table \ref{sources}. We have correlated these 
 sources with the 30 radio sources discovered by Reynoso et al.~(2006) 
 within \rx. We found that the nominal position of one of these radio sources coincides 
 with the peak emission of the X-ray source 11. But cross-correlating all these 18 sources 
 with SIMBAD and NED databases did not result in any identification within a circle 
 of 30 arcsec radius around each source. 

\begin{table}
 \centering
 \caption{Faint X-ray sources detected serendipitously within 
\rx.}\label{sources}
 \begin{tabular}{c c c c c}
 \hline\hline
Source   &     RA (J2000) &  Dec (J2000)  & $\delta$RA & $\delta$Dec\\\hline
         &       h:m:s    &   d:m:s       & arcsec      & arcsec\\ \hline
  1      &  08:52:39.037  & -46:10:50.08  & 0.45 & 0.31\\
  2      &  08:52:12.931  & -46:09:06.68  & 0.37 & 0.25\\
  3      &  08:51:38.185  & -46:10:46.41  & 0.38 & 0.26\\
  4      &  08:52:25.433  & -46:15:57.23  & 0.45 & 0.32\\
  5      &  08:52:22.510  & -46:15:28.37  & 0.50 & 0.48\\
  6      &  08:52:11.708  & -46:14:46.48  & 0.40 & 0.28\\
  7      &  08:52:01.231  & -46:14:18.78  & 0.31 & 0.21\\
  8      &  08:51:04.275  & -46:11:53.72  & 0.45 & 0.32\\
  9      &  08:52:09.804  & -46:22:04.49  & 0.38 & 0.26\\
  10     &  08:51:49.975  & -46:20:41.32  & 0.35 & 0.24\\
  11     &  08:51:01.452  & -46:16:47.83  & 0.30 & 0.23\\
  12     &  08:50:53.378  & -46:16:21.35  & 0.48 & 0.32\\
  13     &  08:50:47.810  & -46:16:09.15  & 0.45 & 0.33\\
  14     &  08:52:04.977  & -46:24:20.52  & 0.44 & 0.30\\
  15     &  08:52:27.994  & -46:26:25.81  & 0.78 & 0.28\\
  16     &  08:52:04.651  & -46:26:28.66  & 0.49 & 0.33\\
  17     &  08:51:43.459  & -46:31:22.22  & 0.50 & 0.28\\
  18     &  08:51:22.030  & -46:30:08.20  & 0.39 & 0.25\\
 \hline
 \end{tabular}
 \end{table}

\subsection{Spectral Analysis}
 As a compromise to select as much as possible events from \cx\ without the 
 contamination from the surrounding remnant emission (see Figure \ref{mos12_image}) 
 we extracted the source spectrum from a circular region of 20 arcsec radius. 
 About 70\% of all point source events are located within this region. Annular 
 regions with radii of $22$ and $35$ arcsec, centered on \cx, were used to extract 
 the background spectra. The background corrected count rates are listed in 
 column 6 of Table 1. Response files were computed for all datasets by using 
 the XMMSAS tasks RMFGEN and ARFGEN. According to the XMMSAS task EPATPLOT, no 
 dataset is affected by CCD pile.

 In order to fit spectra from all three observation runs simultaneously
 we have grouped the energy channels of all spectra dynamically. For the MOS1/2
 data of the April 2001 observation we grouped the data with 
 $\ge$30 counts/bin. For the MOS1/2 and PN data of the May 2003 observation
 we used a grouping with $\ge$100 counts/bin and $\ge$ 165 counts/bin,
 respectively, whereas for the June 2003 observations the MOS1/2 and PN data
 were grouped with $\ge$30 resp.~50 counts/bin.
 The different grouping reflects the differences in photon statistics in
 the various datasets.

 All the spectral fits were performed  in the $0.3-10$ keV energy 
 range by using XSPEC 11.3.1. The spectral models tested include a single blackbody, 
 a double blackbody, a power law, a combination of a blackbody and power law model 
 as well as a broken power law and a thermal bremsstrahlung model. The parameters 
 of all fitted model spectra are summarized in Table \ref{spec_par}. The quoted 
 errors are conservative and are $1\sigma$ for 2 parameters of interest for single 
 component spectral models and for 3 parameters of interest for multi-component 
 models. 

 A single blackbody model gives a parameter set of $T=(4.53\pm 0.05)\times 10^{6} K$, 
 $R=0.29\pm 0.01$ km (for a distance d = 1 kpc)
 and $N_{H}=3.22^{+0.14}_{-0.13}\times 10^{21}$ cm$^{-2}$ 
 ($\chi^{2}_{\nu}=1.21$ for 431 dof). These values are similar to those inferred 
 from Chandra data by Kargaltsev et al.~(2002). However, with a more than 3 times 
 improved photon statistic we find that this simple model cannot describe the 
 data beyond $\sim 3$ keV and hence a single blackbody is not a valid description 
 for the spectrum of \cx. This is in contradiction to the results obtained by 
 Kargaltsev et al.~(2002) using Chandra data only. 

\begin{figure}
 \psfig{file=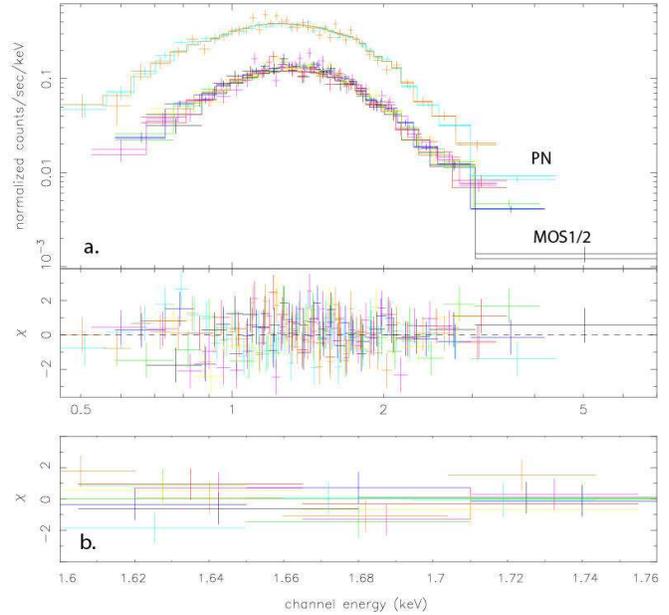,width=8.7cm,clip=}
 \caption[]{
 {\bf a.}\quad Energy spectra of \cx\ observed with the EPIC-PN (upper spectra) and
 the MOS1/2 detectors (lower spectra) with the best-fitting absorbed two component blackbody model 
 ({\it upper panel}), and the contribution to the  $\chi^{2}$ fit statistic ({\it lower panel}). 
 \quad {\bf b.}\quad Details of the $\chi^{2}$ fit distribution of the absorbed two component blackbody model in  
 the $\sim 1.6-1.8$ keV energy range. No line feature is indicated in the fit residuals.} 
  \label{spec}
 \end{figure}

\begin{figure*}
 \centering
 \psfig{file=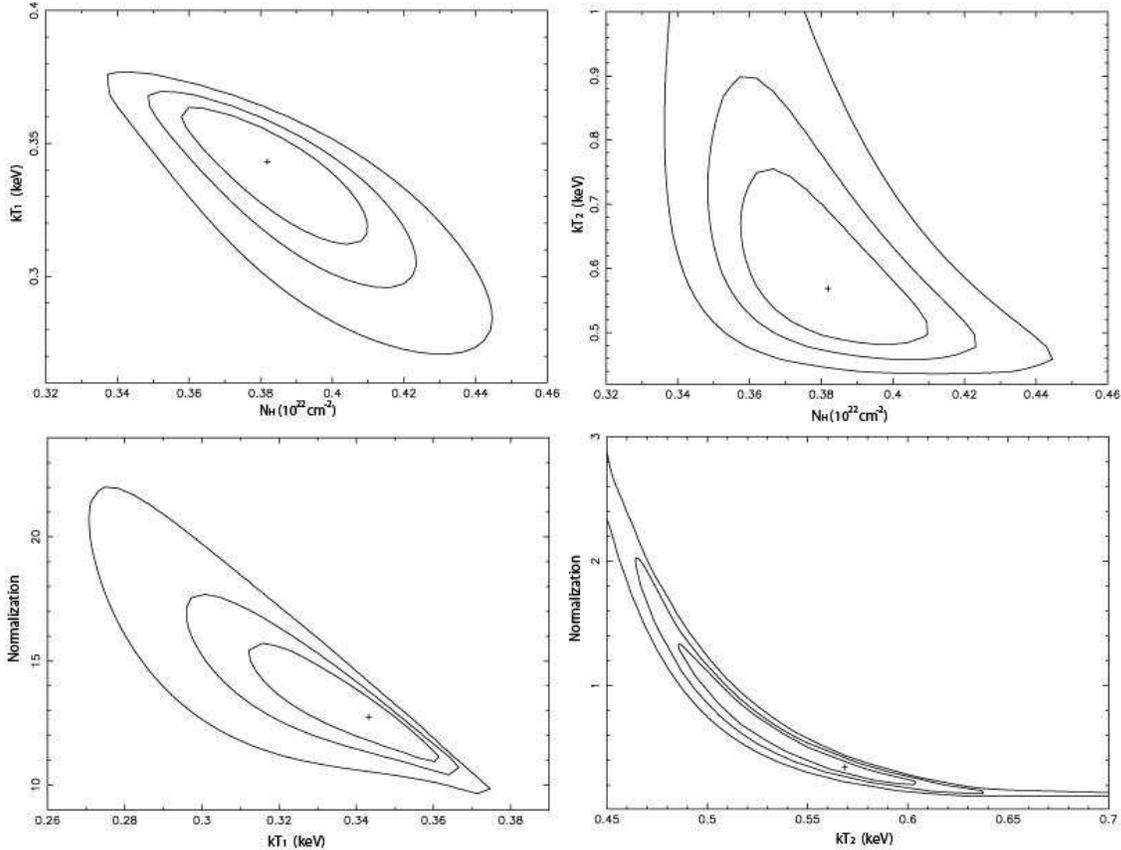,width=15cm,clip=}
 \caption[]{
 $1\sigma$, $2\sigma$ and $3\sigma$ confidence contours for the double blackbody fit
 to the X-ray spectrum of \cx. }
  \label{contour}
  \end{figure*}

 Adding a second thermal component improves the fit and represents very well the observed 
 spectrum up to $\sim 7$ keV (see Figure \ref{spec}). This thermal multi-component 
 model yields a parameter set of $N_{H}=3.82^{+0.36}_{-0.30}\times 10^{21}$ cm$^{-2}$,
 $T_{1}=(2.98^{+0.28}_{-0.47})\times 10^{6} K$, $R_{1}=0.36^{+0.05}_{-0.03}$ km,
 $T_{2}=(6.60^{+3.02}_{-1.18})\times 10^{6} K$ and $R_{2}=59^{+64}_{-43}$ m
 ($\chi^{2}_{\nu}=1.08$ for 429 dof). $R_{1}$ and $R_{2}$ are given for a distance of 1 kpc. 
 The $F-$test statistic indicates
 that the improvement of the goodness-of-fit for adding an extra thermal spectral 
 component is significant at a confidence level of $\ge 99\%$. To further constrain 
 the parameter space for this two component thermal model we calculated the $\chi^{2}_{\nu}$ 
 error contours which are shown in Figure \ref{contour}. We would like to point out 
 that the column density derived by us is consistent with the results from the 
 spectral analysis of \rx\ by Iyudin et al.~(2005). 

 We have also tested whether there is any thermal emission contributed from an emission 
 area compatible with the entire  neutron star surface. However, with $R_{1}$ fixed to
 10 km we found that the goodness-of-fit decreased significantly to $\chi^{2}_{\nu}=1.46$ 
 (for 430 dof).

 Testing non-thermal emission models shows that a single power law model cannot fit 
 the data ($\chi^{2}_{\nu}=2.18$ for 431 dof). However, a model combining a blackbody 
 and a power law yields a goodness-of-fit which is comparable with that obtained with 
 the double blackbody model ($\chi^{2}_{\nu}=1.09$ for 429 dof). The inferred slope of 
 the power law component is $\Gamma$ is $4.21^{+0.30}_{-0.59}$. A similar steep 
 slope has also been observed in the X-ray spectrum of the central compact X-ray source 
 RX J0822-4300 in Puppis-A  (Hui \& Becker 2006). We point out that such a steep 
 power law slope, a priori, cannot be considered to be {\em unphysical} as the nature 
 of the central compact object is still uncertain. Compared with the double blackbody
 model the column density of the combined blackbody plus power law  model,  
 $N_{H}=7.63^{+0.72}_{-1.59}\times10^{21}$ cm$^{-2}$, is higher. However, we find it
 still within the limit of the total Galactic neutral hydrogen column density along
 the direction to \cx\ ($\sim 10^{22}$ cm$^{-2}$; Dickey \& Lockman 1990). The uncertainty 
 in the source distance thus does not support to reject the blackbody plus power law model 
 because of a larger $N_{H}$ value.

 We have also examined whether a thermal brems\-strahlung model, which physically implies
 that the central compact object is surrounded by a hot plasma, yields a viable description. 
 $\chi^{2}_{\nu}$ and $N_{H}$ resulting from this model are similar to the values 
 inferred from the broken power law and hence cannot be considered as an appropriate
 description of the source spectrum (c.f. Table~3).

 From the October 2001 Chandra observation Kargaltsev et al.~(2002) reported a marginal detection
 of a spectral  
 feature at 1.68 keV with a width of  $\sim 100$ eV. The authors found 
 that the shape of this feature is similar to that of an inverse P Cygni profile, suggesting 
 a mass accreting neutron star. However, we have not found any significant 
 feature around 1.68 keV in the spectra obtained by XMM-Newton. Figure \ref{spec}b shows the 
 contribution of the $\chi^{2}$ statistics in the $\sim 1.6-1.8$ keV range for the best-fitted 
 double blackbody model. There is no evidence for any systematic residual in this region. 
 In order to quantify this result we have fitted a Gaussian line of width 100 eV at 1.68 keV. 
 A $3\sigma$ upper limit of such a Gaussian line is 
 $8.27\times 10^{-6}$ photons cm$^{-2}$ s$^{-1}$, or  $2.23\times 10^{-14}$ erg cm$^{-2}$ s$^{-1}$.

\subsection{Timing Analysis}
 The May and June 2003 observations were performed with the EPIC-PN camera operated in
 the small-window mode. The 5.7 ms temporal resolution provided by that mode enables us 
 to search for coherent pulsations in these datasets. For the timing analysis we selected
 all events within $0.3-10$ keV from a circle of 25 arcsec centered on the \cx. This 
 yields 15668 and 5974 events.  
 The background contribution in each observation is about 15\%.

 The barycentric corrections of the arrival times were performed with the XMMSAS task 
 BARYCEN. The position of \cx\ measured by Chandra, RA=$8^{\rm h}52^{\rm m}01^{\rm s}.38$ 
 and Dec=$-46^{\circ}17'53".34$ (J2000), was used for barycentering photon arrival times 
 in both datasets.

 A FFT based period search did not reveal any strong signal of pulsed emission. 
 Taking the candidate periods found by Kargaltsev et al.~(2002), $P=300.8214531$ ms and
 $P=32.92779028$ ms, as initial trials, we carried out a detailed search around these periods 
 by using the $Z^{2}_{m}$ test where $m$ is the numbers of harmonics included 
 (Buccheri et al.~1983). We did not find any promising signals around these test periods 
 either. A pulsed fraction upper limit p was computed using the May 2003 data according to
 
 \begin{center}
 \[
  \quad\quad\quad\quad\quad\quad n_\sigma = \frac{p\,\sqrt{N_{tot}}}{\sqrt{p + \frac{d}{1-d}\,(1-p)}}
 \]
 \end{center}

\noindent
 in which $N_{tot}$ is the total number of detected source counts (12871 cts), $p$ is the pulsed
 fraction (fraction between pulsed and total counts) and $d$ is the duty cycle
 (fraction between pulsed and total time interval). Assuming a duty cycle of\/
 $d=0.5$ (sinusoidal modulation) we compute for $n_\sigma = 3$ a pulsed fraction upper limit
 of $p \sim 3\%$.
 
\section{Discussion}

 We have observed the neutron star \cx\ in the supernova remnant \rx\ with XMM-Newton. 
 The XMM data have invalidated the interpretation that the nature of the diffuse and 
 extended emission surrounding \cx\ is plerionic and powered by the young neutron 
 star. The spectrum of \cx\ is best fitted by a double blackbody spectrum or a blackbody
 plus non-thermal spectral component. The fitted temperatures along with the small size 
 of the emitting regions invalidate the interpretation that the thermal radiation is 
 cooling emission from the entire neutron star surface. If the double black-body model 
 indeed is the right description for the emission scenario of \cx\ it points towards 
 emission from a $\sim 4- 6.5$ MK hot polar-region on the neutron star surface. 

 In the literature various models are proposed to produce hot-spots on the 
 neutron star surface. Almost all magnetospheric emission models predict a 
 bombardment of the polar cap regions by energetic particles accelerated in 
 the magnetosphere backwards to the neutron star surface (cf.~Cheng, Ho \& Ruderman 
 1982; Harding \& Muslimov 2002 and references therein). Though \cx\ is 
 not known to be a radio pulsar it still could be possible that the hot polar 
 caps are heated by particle bombardment while the radio emission itself is
 undetected due to an unfavorable beaming geometry. This is not an unlikely
 scenario. The opening angle of a radio beam is inversely proportional to 
 the square-root of the pulsar's rotation period so that the radio beams
 of slow rotating pulsars can easily miss the observers line of sight and 
 thus keep undetected. 

 The heat transport in neutron stars is accomplished by electrons and 
 positrons. A strong magnetic field thus is expected to have an essential
 impact on the neutron star cooling as it channels the heat 
 along the magnetic field lines and suppresses it perpendicular to it.
 Neutron star cooling with a full treatment of the strong magnetic field
 thus should lead to an anisotropic heat flow and subsequently to an 
 anisotropic surface temperature distribution, with the polar-cap 
 regions getting hotter than the surface temperature of the 
 equatorial regions. This scenario has been modeled recently by 
 Geppert et al.~(2004; 2006). 

 Independent of these scenarios there exists the possibility that \cx\ is
 accreting from fall-back matter or a residual disk and that the
 accreting matter causes the hot-spots on the surface. 

 The low pulsed fraction upper limit is not in disagreement with any of these
 scenarios. When general relativistic effects are taken into account 
 (Page 1995; Hui \& Cheng 2004), pulsations are found to be strongly 
 suppressed and the pulsed fraction is highly dependent on the mass to 
 radius ratio of the star, the orientation of the hot spot and the viewing 
 angle geometry. This is due to the fact that the gravitational bending of 
 light will make more than half of the stellar surface become visible at 
 any instant and hence the contribution of the hot spot will be strongly reduced. 
 If the orientation of the hot spot differs from that of an orthogonal 
 rotator and/or the star has a high mass to radius ratio, then very low 
 amplitude pulsations are expected.

 Anyway, future data to be taken in forthcoming observations  
 will further explore the emission models and hopefully detect the neutron 
 star's rotation period.

\begin{acknowledgements}
The XMM-Newton project is an ESA Science Mission with instruments
and contributions directly funded by ESA Member States and the
USA (NASA). The XMM-Newton project is supported by the
Bundesministerium f\"ur Wirtschaft und Technologie/Deutsches Zentrum
f\"ur Luft- und Raumfahrt (BMWI/DLR, FKZ 50 OX 0001), the Max-Planck
Society and the Heidenhain-Stiftung. 
\end{acknowledgements}

\begin{landscape}
\begin{table}
\centering
\begin{tabular}{llllllll}
\hline
Parameter & BB  & BB+BB & BB+BB                  & PL & BB+PL & BKPL & BREMSS\\
          &     &       & (fix $R_{1}$ at 10 km) &  &  & &\\
	  \hline\hline\\
	  $N_{H}$($10^{21}\mbox{cm}^{-2}$) & $3.215^{+0.135}_{-0.132}$ & $3.819^{+0.356}_{-0.295}$ & $9.212^{+0.313}_{-0.553}$ & $10.994$ & $7.633^{+0.718}_{-1.594}$& $7.422^{+0.421}_{-0.519}$ & $6.821^{+0.146}_{-0.156}$ \\
	  \\
	  $\Gamma_{1}$                     & - & - & - & $4.407$ & $4.208^{+0.298}_{-0.590}$ & $3.050^{+0.170}_{-0.213}$ & -\\
	  \\
	  $\Gamma_{2}$                     & - & - & - & - & - & $5.139^{+0.238}_{-0.188}$ & -\\
	  \\
	  $T_{1}$ ($10^{6}$K)              & $4.533^{+0.048}_{-0.049}$ & $3.983^{+0.281}_{-0.466}$ & $1.358^{+0.022}_{-0.042}$ & - & $4.248^{+0.114}_{-0.147}$ & - & $9.874^{+0.201}_{-0.178}$\\
	  \\
	  $T_{2}$ ($10^{6}$K)              & - & $6.599^{+3.023}_{-1.177}$ & $4.171^{+0.054}_{-0.066}$ & - & - & - & -\\
	  \\
	  $R_{1}$ (km)                     & $0.285^{+0.009}_{-0.008}$ & $0.357^{+0.052}_{-0.034}$ & 10 & - & $0.300^{+0.034}_{-0.021}$ &- & -\\
	  \\
	  $R_{2}$ (km)                     & - & $0.059^{+0.064}_{-0.043}$ & $0.408^{+0.018}_{-0.014}$ & - & - & - & -\\
	  \\
	  $F_{X_{0.5-10keV}}$ (ergs cm$^{-2}$ s$^{-1}$) & $1.905\times10^{-12}$ & $2.113\times10^{-12}$ & $9.916\times10^{-12}$ & $1.720\times10^{-11}$ & $6.324\times10^{-12}$ & $5.426\times10^{-12}$ & $4.162\times10^{-12}$\\
	  \\
	  Reduced $\chi^{2}$              & 1.208 & 1.083 & 1.460 & 2.177 & 1.094 & 1.200 & 1.207\\
	  \\
	  D.O.F                           & 431  & 429 & 430 & 431 & 429 & 429 & 431\\
	  \\
	  \hline
	  \end{tabular}
\caption{Spectral model parameters of \cx\ derived from XMM-Newton data; models: blackbody (BB), two blackbodies 
(BB+BB), power law (PL), blackbody plus power law (BB+PL), broken power law (BKPL), bremsstrahlung (BREMSS).}\label{spec_par}
	  \end{table}
	  \end{landscape}
\end{document}